\documentclass[]{spie}  
\usepackage{algorithm}
\usepackage{booktabs} 
\usepackage{algpseudocode}
\usepackage{tabularx} 
 
\usepackage{amsmath,amsfonts,amssymb}
\usepackage{graphicx}
\usepackage[colorlinks=true, allcolors=blue]{hyperref}

\title{ADP-FL-MedSeg: Adaptive Differential Privacy for Federated Medical Segmentation Across Diverse Modalities}

\author[a]{Puja Saha}
\author[a]{Eranga Ukwatta}
\affil[a]{College of Engineering, University of Guelph, 50 Stone Road E, Guelph, ON, Canada}

\authorinfo{Further author information: (Send correspondence to P.S.)\\P.S.: E-mail: psaha03@uoguelph.ca\\ E.U.: E-mail: eukwatta@uoguelph.ca}

\begin{document} 
\maketitle

\begin{abstract}
Large volumes of medical data remain underutilized because centralizing distributed data is often infeasible due to strict privacy regulations and institutional constraints. In addition, models trained in centralized settings frequently fail to generalize across clinical sites because of heterogeneity in imaging protocols and continuously evolving data distributions arising from differences in scanners, acquisition parameters, and patient populations. Federated learning offers a promising solution by enabling collaborative model training without sharing raw data. However, incorporating differential privacy into federated learning, while essential for privacy guarantees, often leads to degraded accuracy, unstable convergence, and reduced generalization. In this work, we propose an adaptive differentially private federated learning (ADP-FL) framework for medical image segmentation that dynamically adjusts privacy mechanisms to better balance the privacy–utility trade-off. The proposed approach stabilizes training, significantly improves Dice scores and segmentation boundary quality, and maintains rigorous privacy guarantees. We evaluated ADP-FL across diverse imaging modalities and segmentation tasks, including skin lesion segmentation in dermoscopic images, kidney tumor segmentation in 3D CT scans, and brain tumor segmentation in multi-parametric MRI. Compared with conventional federated learning and standard differentially private federated learning, ADP-FL consistently achieves higher accuracy, improved boundary delineation, faster convergence, and greater training stability, with performance approaching that of non-private federated learning under same privacy budgets. These results demonstrate the practical viability of ADP-FL for high-performance, privacy-preserving medical image segmentation in real-world federated settings.
\end{abstract}

\keywords{Optimization,
Adaptive Gradient Clipping,
Decentralized Learning,
HAM10K,
KiTS23,
BraTS24,
Medical Imaging AI,
Privacy-Utility Trade-Off.}

\section{INTRODUCTION}
\label{sec:intro}

Artificial intelligence, particularly deep learning, has demonstrated strong performance in medical imaging tasks such as diagnosis, segmentation, and outcome prediction \cite{holzinger2017we, 10179875, topol2019highperformance}. Despite these advances, translating AI systems into clinical practice remains challenging due to strict regulatory requirements, the high sensitivity of medical data, and substantial heterogeneity in imaging protocols across institutions. Centralized model training requires aggregating large volumes of patient data, which is often constrained by institutional governance processes, storage limitations, consent requirements, and the need for extensive harmonization of multi-site medical images \cite{price2019privacy, mcmahan2017communication}. Furthermore, variations in acquisition settings, annotation quality, scanner hardware, and patient populations reduce the generalization of centrally trained models and necessitate repeated retraining as clinical data distributions evolve \cite{ying2025survey, raza2025meta}.

Federated learning (FL) addresses these challenges by enabling multiple institutions to collaboratively train a shared model without exchanging raw patient data \cite{guan2024survey, daSilva2024survey}. In FL, clients perform local training and communicate only model updates, reducing regulatory risk and eliminating the need for centralized data transfer. Prior work has shown that FL can improve generalization across heterogeneous clinical sites and enable large-scale collaboration in medical imaging applications \cite{rashidi2024potential, tayebi2023enhancing}. Despite these advantages, FL remains vulnerable to privacy leakage through shared gradients or model parameters. Attacks such as gradient inversion and membership inference have demonstrated that sensitive information can be partially recovered from client updates \cite{zhu2019deep, nasr2019comprehensive}. To mitigate these risks, differential privacy (DP) has emerged as a principled approach for providing formal privacy guarantees in FL. Standard DP-FL mechanisms typically rely on per-sample gradient clipping and the addition of fixed, calibrated noise prior to aggregation \cite{dwork2008differential, abadi2016deep}. While effective for privacy protection, this static strategy often leads to degraded performance, particularly for high-dimensional medical image segmentation tasks where gradients exhibit large variance and fine structural details are essential \cite{li2021differentially, chen2022differentially}.

\begin{figure}[H]
    \centering
    \includegraphics[width=0.75\linewidth]{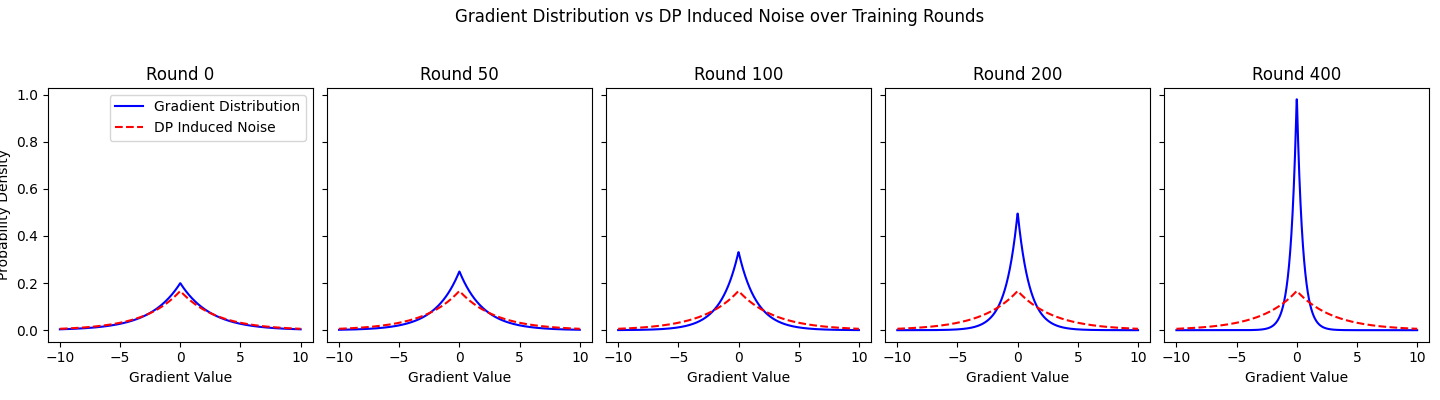}
    \vspace{3pt}
    \caption{Illustrative simulation showing how static differential privacy mechanism fail to adapt with evolving gradient distribution}
    \label{fig:static_dp}
\end{figure}

A key limitation of static DP-FL is that gradient distributions vary across clients and throughout training. Noise levels that are appropriate during early training rounds become excessive as optimization progresses and gradient norms decrease, as illustrated in Figure~\ref{fig:static_dp}. This imbalance reduces the gradient signal-to-noise ratio, slows convergence, and degrades segmentation quality, especially at anatomical boundaries.
\vspace{-6pt}
\vspace{-6pt}
Adaptive differential privacy (ADP) methods aim to address this limitation by dynamically adjusting privacy-related parameters such as clipping thresholds, noise scales, or round-wise privacy allocation. By aligning the injected noise with the evolving gradient distributions of individual clients, ADP approaches can better preserve informative updates and improve training stability. Although adaptive DP strategies have shown promise in improving utility in several learning settings \cite{talaei2024adaptive, zhang2023adaptive, zheng2025sensitivity}, their behavior in federated medical image segmentation remains insufficiently explored. In particular, a noticeable performance gap persists between non-private FL and DP-FL under strict privacy budgets.

In this work, we aim to bridge this gap by proposing adaptive differentially private federated learning framework for medical image segmentation. The proposed approach stabilizes training, reduces noise-induced distortion, and improves generalization under strict privacy guarantees. Extensive experiments across multiple imaging modalities and segmentation tasks demonstrate that our method consistently outperforms standard DP-FL and substantially narrows the performance gap between private and non-private FL.

\section{Literature Review}

Federated learning (FL) has been widely studied as a framework for collaborative model training in medical imaging without centralized data sharing \cite{mcmahan2017communication, rieke2020future, sheller2020federated}. Methods such as FedAvg and its extensions have demonstrated that global optimization is feasible under non-IID data distributions, which are common in multi-institutional clinical settings \cite{mcmahan2017communication}. Prior work has applied FL to a range of medical imaging tasks, including segmentation, classification, and multi-modal analysis, while also highlighting challenges related to client heterogeneity and communication efficiency \cite{rieke2020future, guan2024survey}. Although FL avoids direct exchange of raw patient data, it does not inherently guarantee privacy. Several studies have shown that sensitive information can be inferred from shared gradients or model updates through gradient reconstruction and inversion attacks \cite{zhu2019deep, geiping2020inverting}. Membership inference attacks further demonstrate the risk of identifying whether specific samples participated in training \cite{nasr2019comprehensive}. These findings have motivated the integration of formal privacy-preserving mechanisms into FL frameworks for clinical deployment \cite{rieke2020future}.

Differential privacy (DP) provides mathematically provable privacy guarantees and has become a common approach for securing FL. In DP-FL, clients typically apply per-sample gradient clipping followed by Gaussian noise injection prior to aggregation \cite{dwork2008differential, abadi2016deep}. While effective for privacy protection, multiple studies report substantial performance degradation when DP is applied to high-dimensional medical imaging tasks, particularly segmentation, due to aggressive clipping and noise amplification \cite{li2021differentially, chen2022differentially}. To address this issue, adaptive DP strategies have been proposed that adjust clipping thresholds, noise scales, or privacy allocation during training based on gradient statistics or sensitivity estimates \cite{talaei2024adaptive, zhang2023adaptive, zheng2025sensitivity}. These approaches generally improve the privacy–utility trade-off compared to fixed DP settings; however, most existing methods are designed for centralized learning or low-dimensional tasks. Their effectiveness in federated medical image segmentation, which involves complex spatial structures and heterogeneous client distributions, remains less explored.

Only limited work has examined adaptive DP mechanisms in federated medical imaging tasks that include segmentation. Jiang \textit{et al.} \cite{jiang2024clientleveldifferentialprivacyadaptive} proposed an adaptive intermediary framework evaluated on prostate MRI segmentation and intracranial hemorrhage diagnosis. Their method improves performance over standard DP-FL, but still exhibits a notable gap compared to non-private FL under strict privacy settings. For example, in prostate MRI segmentation, non-private FL achieved Dice scores 87.69±0.12, while the adaptive intermediary recovered performance to 63.28±4.69 at moderate noise levels. This suggests that structural adaptations alone may be insufficient to fully close the privacy–utility gap in high-dimensional federated segmentation.

\section{METHODOLOGY}

We introduce the Adaptive Differentially Private Federated Learning (ADP-FL) framework for medical image segmentation tasks. ADP-FL optimizes the privacy–utility trade-off by synergizing gradient sparsification, adaptive clipping, and dynamic noise injection. 

\begin{figure}[H]
\centering
\includegraphics[width=0.96\linewidth]{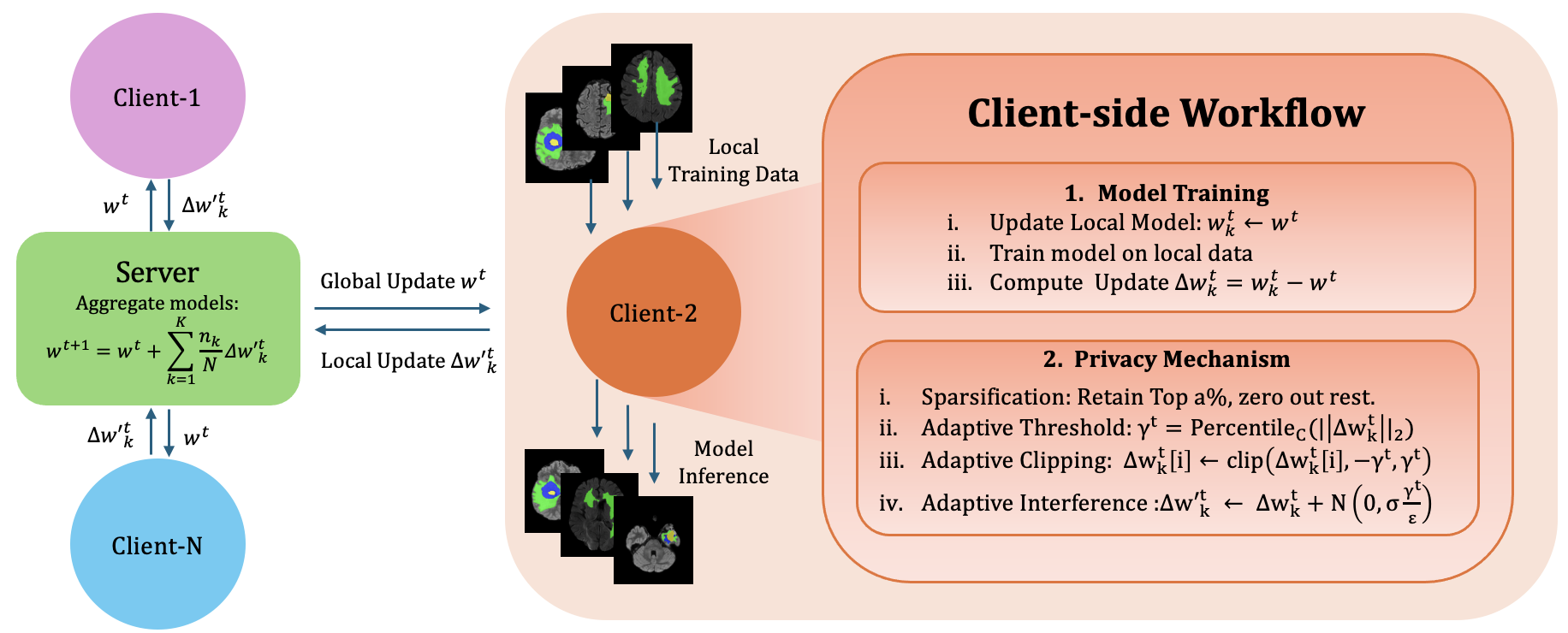}
\caption{Workflow overview of the proposed ADP-FL-MedSeg framework.}
\label{fig:framework}
\end{figure}
\vspace{-3pt}

Consider a federated system consisting of $K$ clients, where each client $k$ maintains a private local dataset $\mathcal{D}_k$ of size $n_k$. At each communication round $t$, the central server broadcasts the global model parameters $w^t$. Each client $k$ performs local training by minimizing the empirical risk:

\begin{equation}
    F_k(w) = \frac{1}{|\mathcal{D}_k|} \sum_{(x_i, y_i) \in \mathcal{D}_k} \ell(w; x_i, y_i)
\end{equation}

where $\ell(\cdot)$ denotes the segmentation loss. Following $E$ local epochs, the client computes the model update:

\begin{equation}
    \Delta w_k^t = w_k^t - w^t
\end{equation}

To mitigate communication overhead, each client applies a sparsification mask, retaining only the top $q\%$ of components. Unlike static approaches, ADP-FL dynamically calibrates the clipping threshold $\gamma_k^t$ based on the observed distribution of the sparsified gradients:

\begin{equation}
    \gamma_k^t = \text{Percentile}_p \big( |\Delta w_k^t| \big)
\end{equation}

The local updates are then projected to ensure the $\ell_2$-norm does not exceed this adaptive threshold, bounding the sensitivity for privacy purposes:

\begin{equation}
    \Delta w_k^t \leftarrow \frac{\Delta w_k^t}{\max\left(1, \frac{\|\Delta w_k^t\|_2}{\gamma_k^t}\right)}
\end{equation}

To provide formal Differential Privacy (DP) guarantees, we employ \textbf{Dynamic Laplace Interference}. By coupling the interference scale to the adaptive threshold $\gamma_k^t$, the framework prevents excessive noise injection as the model converges:

\begin{equation}
    \tilde{w}_k^t = \Delta w_k^t + \text{Laplace}\left(0, \frac{\sigma \gamma_k^t}{\epsilon}\right)
\end{equation}

where $\epsilon$ represents the privacy budget and $\sigma$ denotes the interference scale. Finally, the server performs a weighted aggregation of the sanitized updates:

\begin{equation}
    w^{t+1} = w^t + \sum_{k=1}^{K} \frac{n_k}{N} \tilde{w}_k^t, \quad N = \sum_{k=1}^{K} n_k
\end{equation}
\begin{algorithm}[H]
\caption{One Communication Round of ADP-FL}
\label{alg:adpfl}
\begin{algorithmic}[1]
\Require Global model $w^t$, clients $K$, Epoch $E$, percentile $p$, sparsification $q$, privacy budget $\epsilon$, noise scale $\sigma$
\State Server broadcasts $w^t$ to all clients
\For{each client $k$ in parallel}
    \State Train locally for $E$ epochs to obtain $w_k^t$
    \State Compute update: $\Delta w_k^t = w_k^t - w^t$ \Comment{Equation 2}
    \State Sparsify update by retaining top $q$\% components
    \State Compute adaptive clipping threshold: $\gamma_k^t = \text{Percentile}_p(|\Delta w_k^t|)$ \Comment{Equation 3}
    \State Clip update: $\Delta w_k^t \leftarrow \Delta w_k^t / \max(1, \|\Delta w_k^t\|_2 / \gamma_k^t)$ \Comment{Equation 4}
    \State Inject adaptive noise: $\tilde{w}_k^t = \Delta w_k^t + \text{Laplace}(0, \sigma \gamma_k^t / \epsilon)$ \Comment{Equation 5}
    \State Send $\tilde{w}_k^t$ to server
\EndFor
\State Aggregate client updates: $w^{t+1} = w^t + \sum_{k=1}^{K} \frac{n_k}{N} \tilde{w}_k^t$ \Comment{Equation 6}
\end{algorithmic}
\end{algorithm}

\section{Experimental Setup}

\subsection{Datasets and Preprocessing}
We evaluated the proposed framework across three medical imaging modalities: 2D dermatoscopic images, 3D CT scans, and 3D multi-parametric MRI, using publicly available benchmarks. For each dataset, a centralized hold-out test set was reserved for global evaluation, while the remaining data were partitioned among simulated client sites for federated training.

\textbf{HAM10K:} The Human Against Machine dataset consists of 10,015 dermoscopic images of pigmented skin lesions \cite{ham10000}. We employed a 2D U-Net architecture with an input size of $256 \times 256$. Preprocessing included resizing, pixel intensity normalization to $[0,1]$, and mask binarization for two target classes (lesion and background). For federated training, 500 images were held out for testing, while 9,515 images were distributed across five sites: Site-1 (2,260), Site-2 (2,110), Site-3 (1,940), Site-4 (1,640), and Site-5 (1,565), using a 75\% training and 25\% validation split per site.

\textbf{KiTS23:} The Kidney Tumor Segmentation-2023 challenge dataset contains 489 abdominal 3D CT scans annotated for kidney, cyst, and tumor regions \cite{kits23}. We used the SegResNet architecture with an input volume of $256 \times 256 \times 256$ to segment these three target classes. After reserving 50 cases for testing, 439 scans were distributed among three sites: Site-1 (155), Site-2 (145), and Site-3 (139). Data amount were kept different to emulate clinical setup. Preprocessing involved 3D cropping, resizing, resampling, and intensity rescaling followed by a sigmoid activation.

\textbf{BraTS24:} The Brain Tumor Segmentation-2024 challenge dataset includes 1,621 multi-parametric MRI scans (T1, T1c, T2, FLAIR) \cite{brats24}. A SegResNet architecture with an input size of $176 \times 208 \times 176$ was utilized to segment six target classes: four tumor subregions (NETC, SNFH, ET, RC) and two composite targets (Tumor Core and Whole Tumor). Following the exclusion of 210 test cases, 1,411 scans were distributed across four sites: Site-1 (378), Site-2 (369), Site-3 (356), and Site-4 (308). Preprocessing steps included 3D cropping, resampling, and intensity normalization.

\subsection{Training, Privacy and Implementation}
Clients were trained using the Adam optimizer (learning rate $1 \times 10^{-4}$, weight decay $1 \times 10^{-5}$) with Dice Loss. Each client performed 1 local epoch per communication round for a total of 400 rounds. Model updates were aggregated using data-weighted averaging. Dropout (0.2) and cosine annealing were applied to stabilize the learning process. To ensure privacy, gradients were sparsified (top 90\% retention) and perturbed via Adaptive Differential Privacy (ADP) using Laplacian noise ($\epsilon=0.001, \sigma=1$), with clipping thresholds set at the 95\textsuperscript{th} percentile of gradient norms. Due to DP-induced fluctuations, we maintained a dual-model saving strategy, independently tracking the best-performing and the most recent global models. Performance was measured via the Dice Similarity Coefficient (DSC). All experiments were repeated three times to report global mean $\pm$ standard deviation, while the best overall run was used to calculate the patient-level $mean \pm std$ across individual cases. All experiments were implemented using Python 3.12, NVFlare for federated orchestration, PyTorch, and MONAI. Training was conducted on Linux-operated Compute Canada clusters equipped with NVIDIA H100 GPUs.

\section{Results and Discussion}
This section presents comprehensive experimental results and discussion on three different medical image segmentation tasks:  skin lesion segmentation (HAM10K), kidney tumor segmentation (KiTS23), and multi-subregion brain tumor segmentation (BraTS24). Results are organized into two parts: (i) comparison with baseline methods and (ii) hyperparameter sensitivity analysis. All evaluations are performed on centrally held-out test sets, incorporating quantitative metrics, convergence analyses, and qualitative visual assessments.

\subsection{Comparison with Baseline Methods}
We compare the proposed Adaptive Differentially Private Federated Learning (ADP-FL) framework with Non-Private Federated Learning (NP-FL) and standard Differentially Private Federated Learning (DP-FL) to quantify performance degradation due to privacy constraints and to evaluate how effectively ADP-FL mitigates this trade-off at same privacy budget.

\subsubsection{HAM10K}  
The performance metrics for NP-FL, DP-FL, and the proposed ADP-FL are summarized in Table~\ref{tab:ham_dsc}, reporting both inter-run stability (mean ± std across three runs) and patient-level reliability (mean ± std across patients). The results show that ADP-FL recovers nearly all performance lost under standard DP-FL and, at the patient level, even slightly exceeds the non-private baseline.

\begin{table}[H]
\centering
\caption{Comparative performance (DSC \%) for NP-FL, DP-FL, and ADP-FL. Results are reported as \textit{mean $\pm$ std across three runs / mean $\pm$ std across patients for the best model update.}}
\vspace{3pt}
\begin{tabular}{lccc}
\toprule
\textbf{Method} & \textbf{NP-FL (N=500)} & \textbf{DP-FL (N=500)} & \textbf{ADP-FL (N=500)} \\
\midrule
DSC (\%) &
93.12 ± 0.04 / 95.18 ± 6.51 &
85.38 ± 0.04 / 86.88 ± 14.38 &
92.81 ± 0.06 / 96.27 ± 6.93 \\
\bottomrule
\end{tabular}
\label{tab:ham_dsc}
\end{table}
\vspace{-5pt}

\begin{figure}[H]
\centering
\begin{tabular}{cc}
\includegraphics[width=0.4\textwidth]{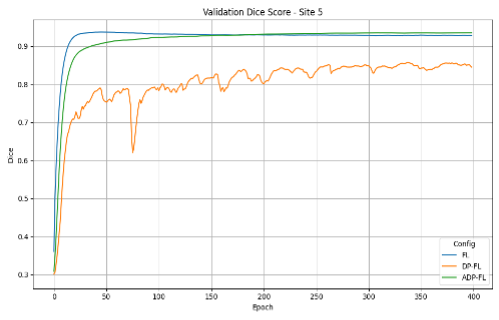} &
\includegraphics[width=0.55\textwidth]{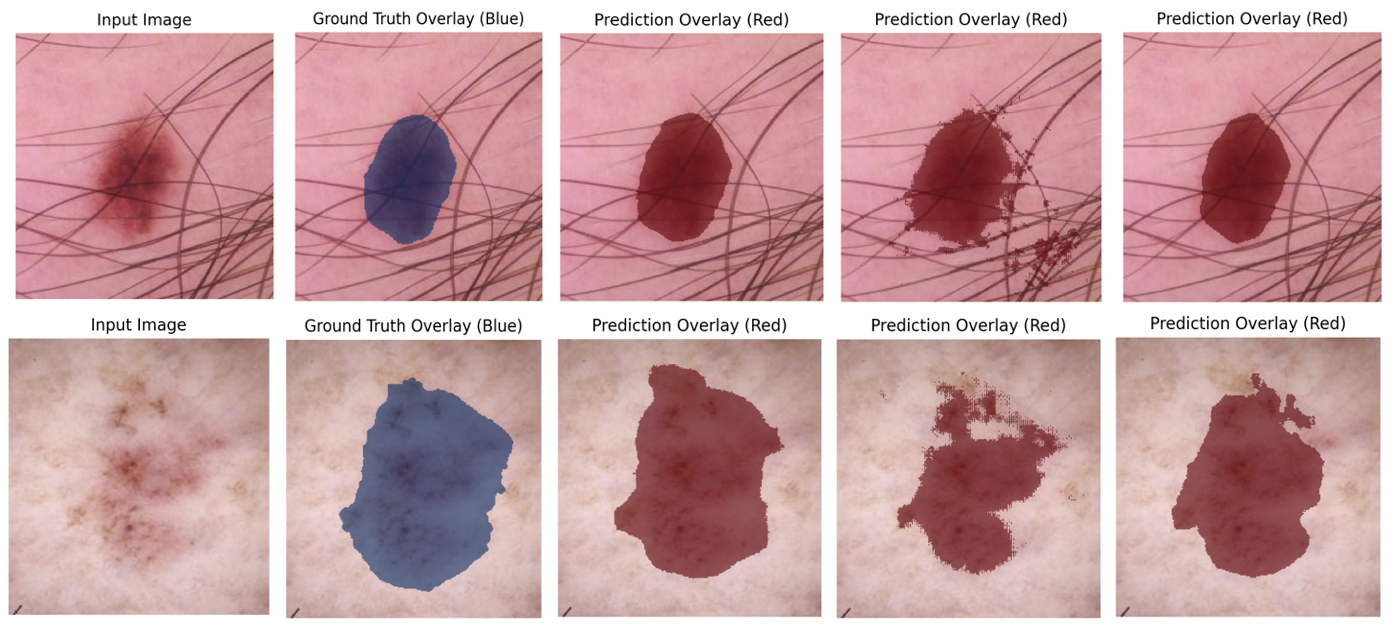}
\end{tabular}
\caption{HAM10K results: (Left) DSC convergence trajectory for site-5. (Right) Qualitative segmentation comparisons between methods.}
\label{fig:ham_results}
\end{figure}

Crucially, ADP-FL maintains high consistency; its patient-level standard deviation ($6.93\%$) is remarkably close to that of NP-FL ($6.51\%$), whereas DP-FL exhibits much higher variability ($14.38\%$). This indicates that while standard DP-FL leads to inconsistent segmentation across different cases, ADP-FL provides a robust and reliable model for diverse patient data. As illustrated in Figure~\ref{fig:ham_results}, the convergence analysis at site-5 confirms that ADP-FL stabilizes the training process and closely tracks the NP-FL trajectory, whereas DP-FL remains noisy and delayed. The qualitative masks (Figure~\ref{fig:ham_results}, right) further validate these findings; ADP-FL preserves sharp lesion boundaries and fine structural details that are often blurred or lost under the rigid gradient clipping of standard DP-FL. By replacing static constraints with an adaptive clipping and noise mechanism, ADP-FL successfully mitigates the primary sources of accuracy degradation in private federated learning.

\vspace{-5pt}

\subsubsection{KiTS23}

Table~\ref{tab:kits_dsc} summarizes NP-FL, DP-FL, and ADP-FL segmentation performance on KiTS23 across Kidney (KTC), Tumor+Cyst, and Tumor regions, reporting both inter-run stability and patient-level reliability on 50 test cases. Standard DP-FL shows substantial degradation in tumor-related accuracy, with mean DSC dropping to $69.08\%$ for Tumor+Cyst and $66.29\%$ for Tumor. ADP-FL mitigates this loss, improving DSC by $8.18$ and $6.55$ points across runs, and by over $10$ points at the patient level, bringing performance closer to the non-private baseline while reducing case-to-case variability. These improvements suggest that adaptive clipping better preserves task-relevant gradients in critical tumor regions and complex 3D structures. Convergence analysis and qualitative visualizations (Figure-4) further demonstrate smoother optimization dynamics and more coherent recovery of tumor and cystic structures under ADP-FL.

\begin{table}[H]
\centering
\caption{KiTS23 Dice Similarity Coefficient (DSC \%). Results are reported as \textit{Mean across three runs / Patient-level Mean for the best run}.}
\label{tab:kits_dsc}
\vspace{3pt}
\resizebox{\textwidth}{!}{%
\begin{tabular}{lccc}
\toprule
\textbf{Method} & \textbf{KTC (N=50)} & \textbf{Tumor+Cyst (N=50)} & \textbf{Tumor (N=50)} \\
\midrule
NP-FL  & 94.92 ± 0.36 / 95.22 ± 1.92 & 81.00 ± 1.03 / 81.93 ± 17.60 & 79.79 ± 1.06 / 80.94 ± 18.18 \\
DP-FL  & 91.94 ± 0.27 / 91.72 ± 5.89 & 69.08 ± 0.98 / 68.00 ± 23.85 & 66.29 ± 1.09 / 65.34 ± 26.38 \\
ADP-FL & 93.63 ± 0.60 / 93.96 ± 2.80 & 77.26 ± 0.60 / 78.96 ± 19.95 & 72.84 ± 2.25 / 76.09 ± 22.71 \\
\bottomrule
\end{tabular}%
}
\end{table}
\vspace{-10pt}

\vspace{-5pt}
\begin{figure}[H]
\centering
\begin{tabular}{cc}
\includegraphics[width=0.33\textwidth]{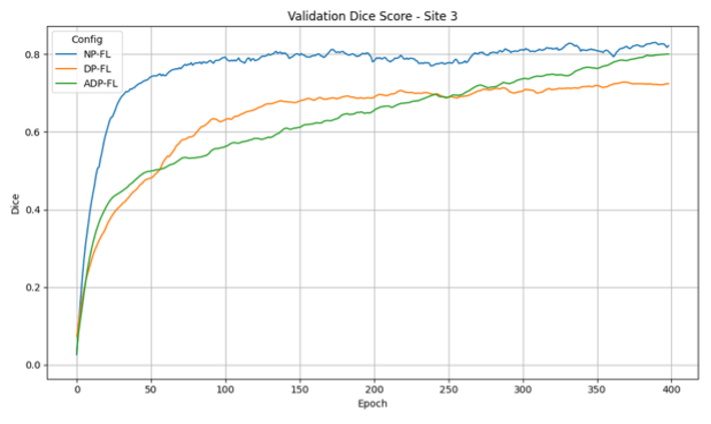} &
\includegraphics[width=0.63\textwidth]{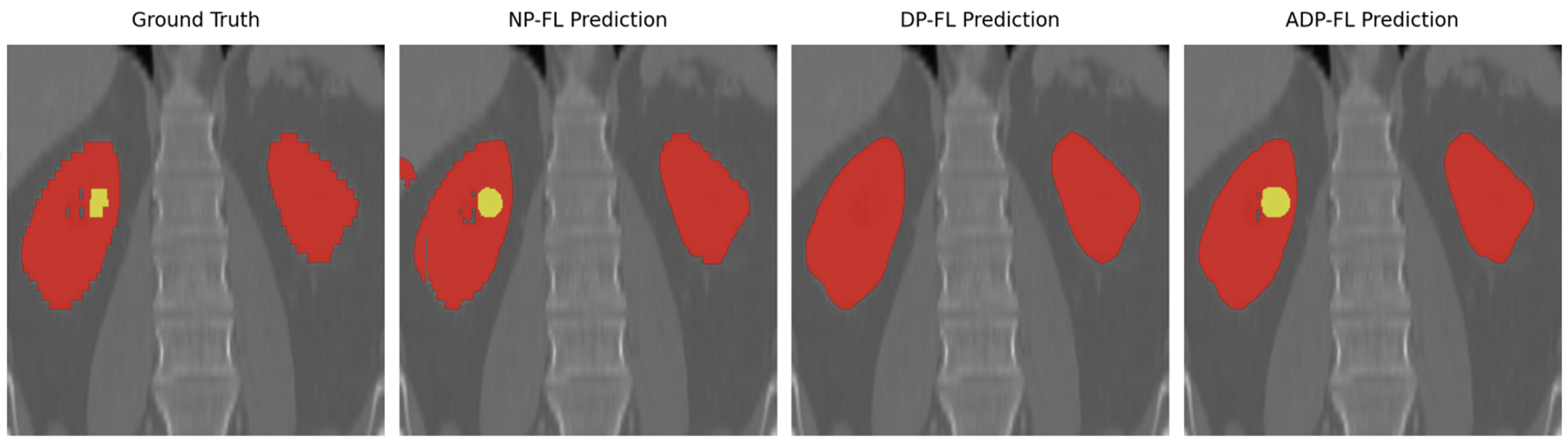}
\end{tabular}
\caption{KiTS23: (Left) Convergence graph for site-3. (Right) Qualitative segmentation masks for Patient case 00581.}
\label{fig:kits_conv_qual}
\end{figure}
\vspace{-9pt}

\subsubsection{BraTS24}

Table~\ref{tab:brats_dsc_runs} summarizes NP-FL, DP-FL, and ADP-FL segmentation performance on BraTS24 across six tumor subregions, reporting mean DSC $\pm$ std across three runs. Standard DP-FL suffers severe degradation in low-contrast and heterogeneous regions such as NETC and ET, with mean DSC dropping below $26\%$ and $57\%$, respectively. In contrast, ADP-FL substantially narrows this gap, achieving mean DSC of $42.05\%$ for NETC and $72.27\%$ for ET, while also improving performance across all other subregions. Patient-level results (Table~\ref{tab:brats_dsc_patients}) highlight that DP-FL exhibits high variability across cases, particularly in challenging subregions, whereas ADP-FL reduces this variance and maintains mean DSC values much closer to the non-private baseline. These improvements indicate that adaptive clipping preserves task-relevant gradient information for complex multi-subregion segmentation under privacy constraints. Convergence behavior and qualitative visualizations (Figure-5) further demonstrate smoother optimization and more anatomically coherent tumor core predictions under ADP-FL, compared to the fragmented and unstable segmentations observed with rigid DP-FL.

\begin{table}[H]
\centering
\caption{BraTS24 DSC (mean ± std) across three runs for NP-FL, DP-FL, and ADP-FL.}
\vspace{3pt}
\resizebox{\textwidth}{!}{
\begin{tabular}{lcccccc}
\toprule
\textbf{Method} & \textbf{ET (N=165)} & \textbf{NETC (N=92)} & \textbf{SNFH (N=210)} & \textbf{RC (N=179)} & \textbf{TC (N=165)} & \textbf{WT (N=210)} \\
\midrule
NP-FL  & 77.56 ± 0.21 & 57.64 ± 0.56 & 87.28 ± 0.06 & 76.86 ± 0.74 & 76.46 ± 0.19 & 88.62 ± 0.06 \\
DP-FL  & 56.77 ± 0.33 & 24.69 ± 0.78 & 72.89 ± 0.40 & 55.01 ± 0.07 & 56.22 ± 0.77 & 76.72 ± 0.48 \\
ADP-FL & 72.27 ± 0.22 & 42.05 ± 4.60 & 85.18 ± 0.13 & 72.04 ± 0.22 & 70.97 ± 0.24 & 86.78 ± 0.06 \\
\bottomrule
\end{tabular}%
}
\label{tab:brats_dsc_runs}
\end{table}

\vspace{-10pt}
\begin{table}[H]
\centering
\caption{BraTS24 DSC (mean ± std) across patients for the best run for NP-FL, DP-FL, and ADP-FL.}
\vspace{3pt}
\resizebox{\textwidth}{!}{%
\begin{tabular}{lcccccc}
\toprule
\textbf{Method} & \textbf{ET (N=165)} & \textbf{NETC (N=92)} & \textbf{SNFH (N=210)} & \textbf{RC (N=179)} & \textbf{TC (N=165)} & \textbf{WT (N=210)} \\
\midrule
NP-FL  & 77.85 ± 23.04 & 58.38 ± 29.41 & 87.36 ± 13.90 & 77.47 ± 24.97 & 76.72 ± 23.40 & 88.71 ± 12.22 \\
DP-FL  & 57.51 ± 26.64 & 25.97 ± 24.69 & 73.70 ± 18.68 & 56.10 ± 29.86 & 57.88 ± 26.80 & 76.48 ± 17.02 \\
ADP-FL & 72.06 ± 24.75 & 46.27 ± 29.50 & 85.35 ± 15.78 & 72.31 ± 27.03 & 71.29 ± 24.57 & 86.86 ± 13.96 \\
\bottomrule
\end{tabular}%
}
\label{tab:brats_dsc_patients}
\end{table}
\vspace{-4pt}

\vspace{-12pt}
\begin{figure}[H]
\centering
\begin{tabular}{cc}
\includegraphics[width=0.33\textwidth]{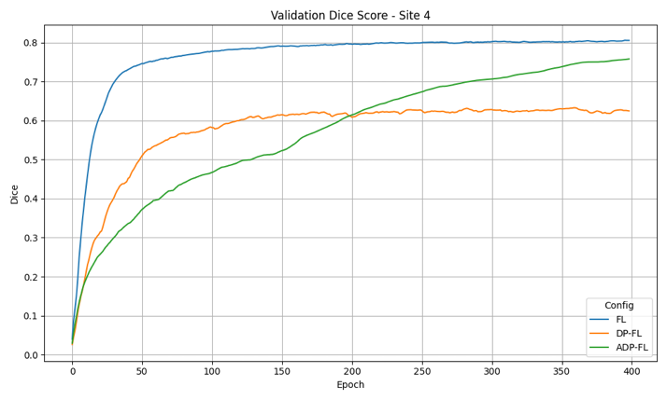} &
\includegraphics[width=0.62\textwidth]{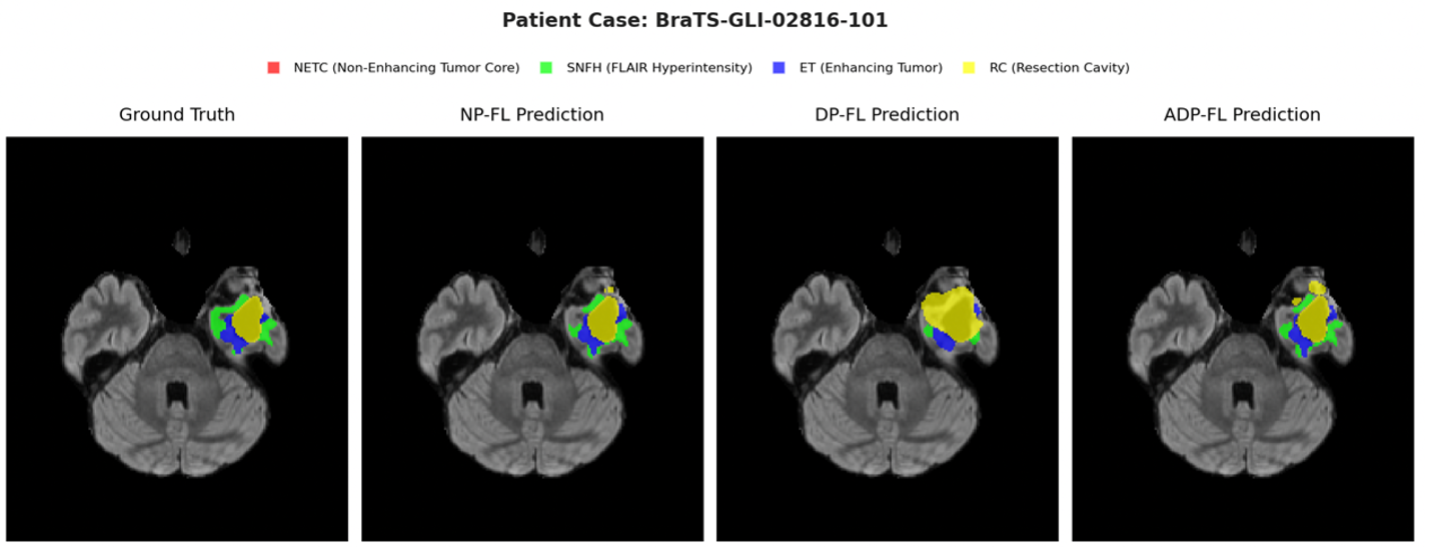}
\end{tabular}
\caption{BraTS24: (Left) Convergence graph of site-4. (Right) Qualitative segmentation masks.}
\label{fig:brats_conv_qual}
\end{figure}
\vspace{-6pt}

\subsection{Hyperparameter Sensitivity Analysis}

We analyze the sensitivity of ADP-FL to the clipping percentile ($p$), which is introduced as an additional hyperparameter governing the degree of gradient truncation and interference. Experiments are conducted with $p$ ranging from the 70th to the 95th percentile to examine the trade-off between gradient information retention and noise injection. From a theoretical perspective, lower values of $p$ lead to excessive truncation of informative updates, whereas higher values increase the clipping threshold $\gamma_k^t$ and consequently amplify the magnitude of noise required to maintain differential privacy. This analysis assesses whether a single clipping percentile can generalize across datasets of varying complexity, including HAM10K, KiTS23, and BraTS24, or whether task-specific tuning is necessary to get optimal performance. 

\subsubsection{HAM10K}
Table~\ref{tab:ham_clip} summarizes segmentation performance on the HAM10K dataset across different clipping percentiles, while Figure-6 illustrates the corresponding convergence behavior and qualitative segmentation results. Although convergence trajectories and predicted masks exhibit nearly identical optimization dynamics and visual consistency across all tested configurations, quantitative performance shows a gradual improvement as the clipping percentile increases from the 70\textsuperscript{th} to the 95\textsuperscript{th} percentile, reaching a maximum Dice Similarity Coefficient (DSC) of $92.81\%$. Importantly, performance variability remains minimal at both the inter-run and patient levels, indicating that ADP-FL is highly robust to the choice of clipping percentile for this 2D dermoscopic segmentation task.

\begin{table}[H]
\centering
\caption{Impact of clipping percentiles on HAM10K DSC performance. Results highlight inter-run stability and intra-patient reliability.}
\label{tab:ham_clip}
\vspace{6pt}
\resizebox{\linewidth}{!}{%
\begin{tabular}{lcccccc}
\toprule
\textbf{Metric (\%)} & \textbf{70\textsuperscript{th}} & \textbf{75\textsuperscript{th}} & \textbf{80\textsuperscript{th}} & \textbf{85\textsuperscript{th}} & \textbf{90\textsuperscript{th}} & \textbf{95\textsuperscript{th}} \\
\midrule
Mean Across Runs     & $92.37 \pm 0.14$ & $92.57 \pm 0.20$ & $92.71 \pm 0.14$ & $92.61 \pm 0.02$ & $92.61 \pm 0.08$ & $92.81 \pm 0.06$ \\
Mean Across Patients & $95.67 \pm 8.11$ & $96.06 \pm 7.62$ & $95.99 \pm 7.87$ & $96.24 \pm 7.37$ & $96.26 \pm 7.29$ & $96.27 \pm 6.93$ \\
\bottomrule
\end{tabular}}
\end{table}
\vspace{-6pt}

\vspace{-6pt}
\begin{figure}[H]
\centering
\begin{tabular}{cc}
\includegraphics[width=0.42\linewidth]{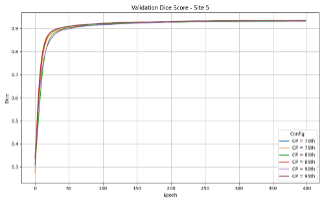} &
\includegraphics[width=0.52\linewidth]{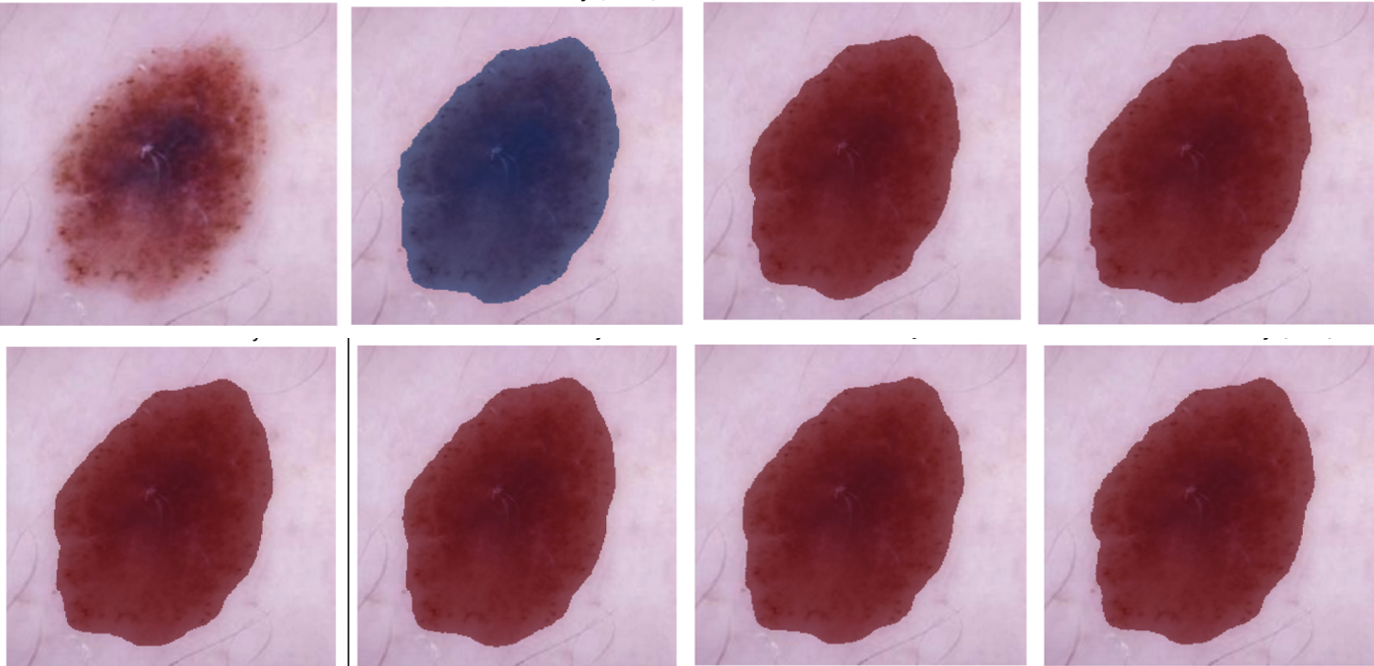}
\end{tabular}
\caption{HAM10K sensitivity analysis: (Left) DSC convergence trajectories for various percentiles at site-5. (Right) Qualitative comparison demonstrating visual invariance to percentile shifts.}
\label{fig:ham_conv_qual}
\end{figure}
\vspace{-5pt}

\subsubsection{KiTS23} 
Sensitivity to clipping is more pronounced on KiTS23 due to heterogeneous tumor morphology and the inherently challenging 3D nature of the segmentation task. As reported in Tables~\ref{tab:kits_clip_runs} and~\ref{tab:kits_clip_patients}, aggressive clipping at lower percentiles substantially degrades performance for the Tumor and Tumor+Cyst regions, increasing variability across runs. Performance improves markedly at the 80th and 85th percentiles, with the 85th percentile achieving the highest overall DSC across tumor-related classes. The 95th percentile remains stable and competitive, indicating that ADP-FL maintains robust performance across a reasonably wide range of higher clipping percentiles. Figure-7 further illustrates the convergence behavior and qualitative segmentations, highlighting that higher percentiles better preserve task-relevant gradients, leading to improved accuracy in complex 3D tumor structures.

\vspace{-6pt}
\begin{table}[H]
\centering
\caption{KiTS23 DSC (mean ± std) across three runs for different clipping percentiles.}
\label{tab:kits_clip_runs}
\vspace{3pt}
\resizebox{\textwidth}{!}{%
\begin{tabular}{lcccccc}
\toprule
\textbf{Class} & \textbf{70\textsuperscript{th}} & \textbf{75\textsuperscript{th}} & \textbf{80\textsuperscript{th}} & \textbf{85\textsuperscript{th}} & \textbf{90\textsuperscript{th}} & \textbf{95\textsuperscript{th}} \\
\midrule
KTC\textsuperscript{*} & 94.51 ± 0.40 & 93.64 ± 1.48 & 93.27 ± 1.32 & 93.97 ± 0.38 & 93.69 ± 0.27 & 93.63 ± 0.60 \\
Tumor+Cyst             & 60.64 ± 18.51 & 66.82 ± 8.72 & 70.75 ± 6.21 & 77.39 ± 0.49 & 77.36 ± 1.40 & 77.26 ± 0.60 \\
Tumor                  & 66.89 ± 4.25 & 64.55 ± 5.05 & 70.39 ± 5.64 & 75.28 ± 1.02 & 74.62 ± 1.31 & 72.84 ± 2.25 \\
\bottomrule
\addlinespace
\multicolumn{7}{l}{\textsuperscript{*}KTC: Kidney + Tumor + Cyst} \\
\end{tabular}%
}
\end{table}
\vspace{-10pt}

\vspace{-6pt}
\begin{table}[H]
\centering
\caption{KiTS23 patient-level DSC (mean ± std) for the best run across clipping percentiles.}
\label{tab:kits_clip_patients}
\vspace{3pt}
\resizebox{\textwidth}{!}{%
\begin{tabular}{lcccccc}
\toprule
\textbf{Class} & \textbf{70\textsuperscript{th}} & \textbf{75\textsuperscript{th}} & \textbf{80\textsuperscript{th}} & \textbf{85\textsuperscript{th}} & \textbf{90\textsuperscript{th}} & \textbf{95\textsuperscript{th}} \\
\midrule
KTC\textsuperscript{*} & 92.15 ± 6.19  & 91.63 ± 6.18  & 95.02 ± 2.07  & 94.65 ± 2.78  & 94.78 ± 2.52  & 93.96 ± 2.80  \\
Tumor + Cyst           & 70.21 ± 20.82 & 68.35 ± 22.82 & 74.23 ± 26.07 & 64.36 ± 30.89 & 77.27 ± 22.51 & 78.96 ± 19.95 \\
Tumor                  & 67.49 ± 23.36 & 65.33 ± 24.98 & 72.13 ± 27.53 & 63.07 ± 31.01 & 75.25 ± 24.61 & 76.09 ± 22.71 \\
\bottomrule
\addlinespace
\multicolumn{7}{l}{\textsuperscript{*}KTC: Kidney + Tumor + Cyst} \\
\end{tabular}%
}
\end{table}
\vspace{-10pt}

\begin{figure}[H]
\centering
\begin{tabular}{cc}
\includegraphics[width=0.32\linewidth]{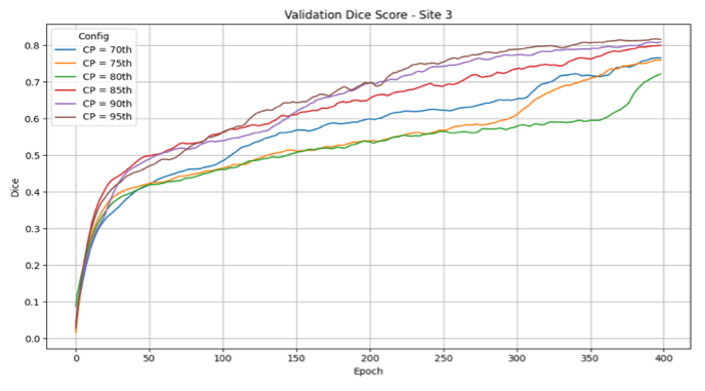} &
\includegraphics[width=0.64\linewidth]{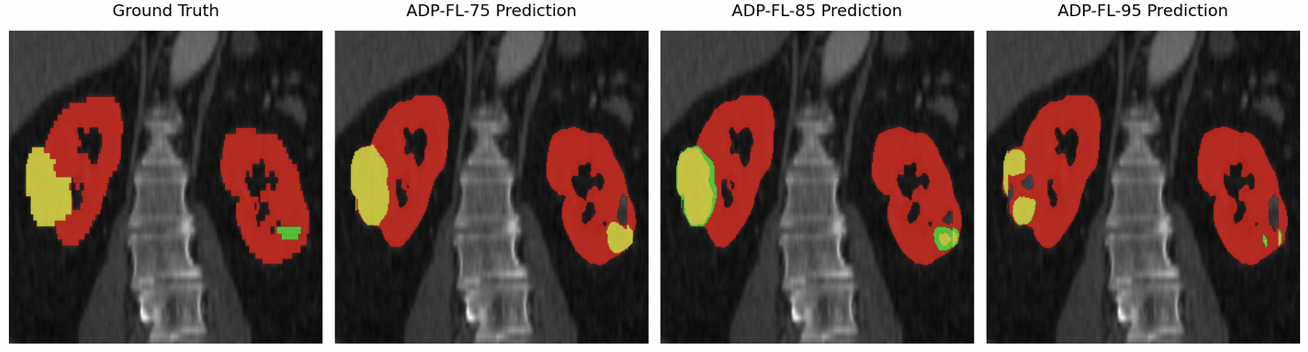}
\end{tabular}
\caption{KiTS23 sensitivity analysis: (Left) DSC convergence trajectories for various percentiles at site-3. (Right) Qualitative comparison demonstrating visual invariance to percentile shifts.}
\label{fig:ham_conv_qual}
\end{figure}
\vspace{-5pt}

\subsubsection{BraTS24} 
BraTS24 demonstrates the highest sensitivity to gradient clipping, which can be attributed to the combination of multi-class targets, multi-modal inputs, and pronounced class imbalance across tumor subregions. As shown in Tables~\ref{tab:brats_clip_runs} and~\ref{tab:brats_clip_patients}, restrictive clipping at lower percentiles substantially suppresses informative gradients, particularly for the NETC and TC classes, resulting in notable degradation of segmentation performance and increased variability across runs and patients. As the clipping percentile increases, performance improves consistently across most classes, indicating more effective preservation of task-relevant gradient information. The highest overall DSC is achieved at the 95th percentile, while the 90th percentile occasionally yields comparable performance for select classes but with reduced robustness. Figure~\ref{fig:brats_conv_qual} further illustrates the associated convergence behavior and qualitative segmentation outcomes, confirming that higher clipping thresholds lead to more stable optimization and more accurate delineation of complex tumor structures in this challenging multi-class setting.

\begin{table}[H]
\centering
\caption{BraTS24 DSC (mean ± std) across three runs for different clipping percentiles.}
\label{tab:brats_clip_runs}
\vspace{3pt}
\resizebox{\textwidth}{!}{%
\begin{tabular}{lcccccc}
\toprule
\textbf{Class} & \textbf{70\textsuperscript{th}} & \textbf{75\textsuperscript{th}} & \textbf{80\textsuperscript{th}} & \textbf{85\textsuperscript{th}} & \textbf{90\textsuperscript{th}} & \textbf{95\textsuperscript{th}} \\
\midrule
ET   & 47.43 ± 5.63  & 44.59 ± 5.47  & 55.82 ± 13.40 & 66.79 ± 2.92  & 69.19 ± 1.44  & 72.27 ± 0.22 \\
NETC & 18.23 ± 1.28  & 10.69 ± 5.72  & 26.32 ± 11.43 & 26.54 ± 11.62 & 38.24 ± 9.04  & 42.05 ± 4.60 \\
SNFH & 75.77 ± 9.95  & 81.44 ± 2.91  & 85.02 ± 0.63  & 84.96 ± 0.84  & 85.03 ± 0.27  & 85.18 ± 0.13 \\
RC   & 53.20 ± 7.07  & 59.29 ± 8.98  & 59.54 ± 9.09  & 60.87 ± 12.12 & 67.63 ± 3.17  & 72.04 ± 0.22 \\
TC   & 45.51 ± 13.98 & 36.69 ± 21.02 & 55.54 ± 7.53  & 67.98 ± 0.74  & 67.72 ± 1.25  & 70.97 ± 0.24 \\
WT   & 83.16 ± 3.28  & 86.12 ± 0.42  & 86.41 ± 0.32  & 86.39 ± 0.28  & 86.63 ± 0.18  & 86.78 ± 0.06 \\
\bottomrule
\end{tabular}%
}
\end{table}
\vspace{-10pt}

\vspace{-6pt}
\begin{table}[H]
\centering
\caption{BraTS24 DSC (mean ± std) across patients for different clipping percentiles.}
\label{tab:brats_clip_patients}
\vspace{3pt}
\resizebox{\textwidth}{!}{%
\begin{tabular}{lcccccc}
\toprule
\textbf{Class} & \textbf{70\textsuperscript{th}} & \textbf{75\textsuperscript{th}} & \textbf{80\textsuperscript{th}} & \textbf{85\textsuperscript{th}} & \textbf{90\textsuperscript{th}} & \textbf{95\textsuperscript{th}} \\
\midrule
ET   & 57.03 ± 26.15 & 55.59 ± 25.27 & 68.04 ± 25.96 & 63.18 ± 26.55 & 47.52 ± 27.50 & 72.06 ± 24.75 \\
NETC & 25.29 ± 24.74 & 25.35 ± 24.86 & 16.96 ± 20.30 & 32.66 ± 27.67 & 33.84 ± 26.64 & 46.27 ± 29.50 \\
SNFH & 73.85 ± 18.25 & 72.62 ± 18.11 & 85.71 ± 15.54 & 85.55 ± 15.05 & 84.14 ± 17.24 & 85.35 ± 15.78 \\
RC   & 56.33 ± 29.49 & 55.72 ± 29.44 & 70.47 ± 28.26 & 66.24 ± 29.60 & 68.07 ± 28.87 & 72.31 ± 27.03 \\
TC   & 57.88 ± 27.22 & 56.86 ± 27.05 & 68.64 ± 26.55 & 68.69 ± 25.67 & 66.40 ± 27.09 & 71.29 ± 24.57 \\
WT   & 77.29 ± 15.68 & 76.11 ± 15.73 & 86.57 ± 14.34 & 86.66 ± 13.77 & 86.64 ± 14.66 & 86.86 ± 13.96 \\
\bottomrule
\end{tabular}%
}
\end{table}
\vspace{-10pt}

\vspace{-6pt}
\begin{figure}[H]
\centering
\begin{tabular}{cc}
\includegraphics[width=0.38\linewidth]{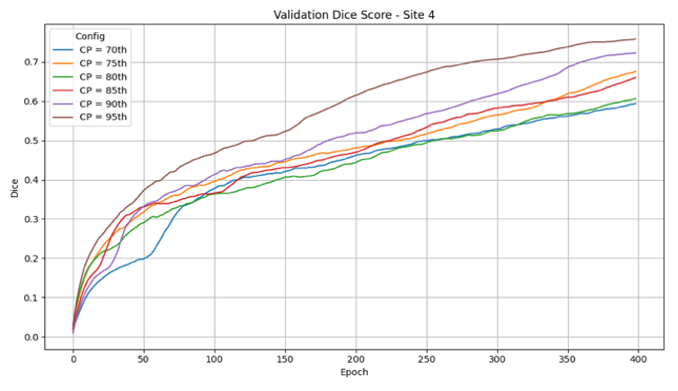} &
\includegraphics[width=0.58\linewidth]{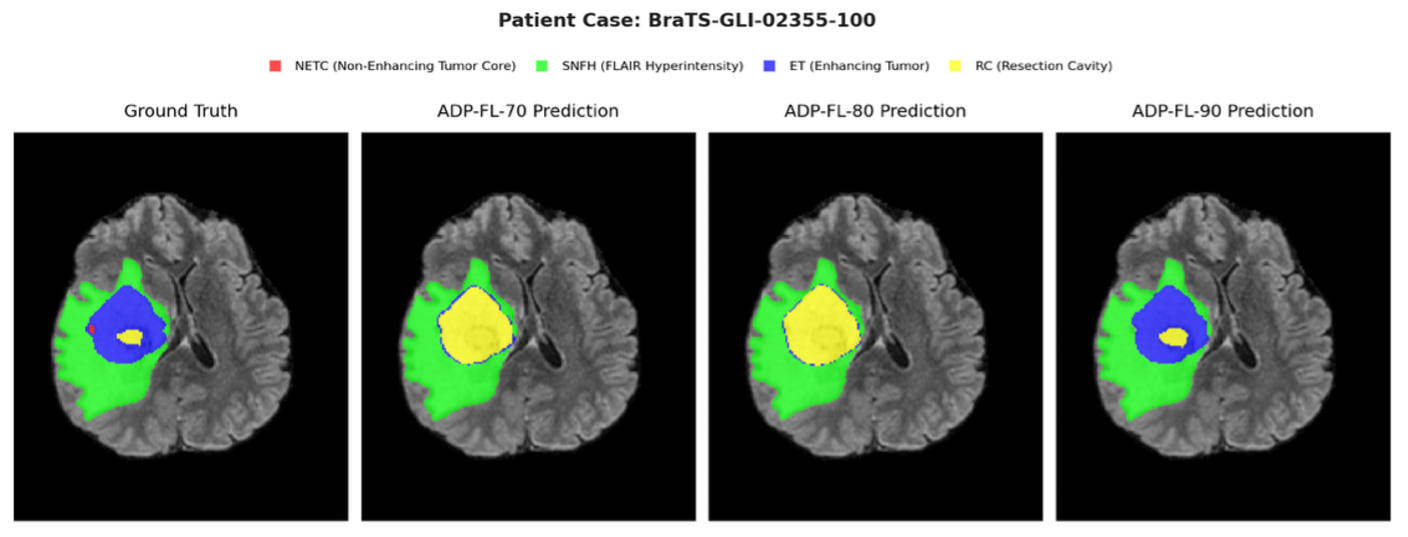}
\end{tabular}
\caption{BraTS24 sensitivity analysis: (Left) convergence graph of site-4. (Right) Qualitative results.}
\label{fig:brats_conv_qual}
\end{figure}
\vspace{-15pt}

\subsection{Discussion}
Across all datasets, ADP-FL consistently outperforms standard DP-FL while approaching the accuracy of non-private FL (NP-FL), demonstrating a favorable privacy–utility trade-off across diverse segmentation settings. On simpler tasks such as HAM10K, ADP-FL recovers nearly all performance lost under DP-FL, while maintaining high inter-run stability and strong patient-level consistency. In these settings, adaptive clipping effectively preserves informative gradients, enabling the model to retain fine structural details in qualitative segmentation masks and achieve convergence behavior that closely mirrors NP-FL. For more complex and heterogeneous tasks, including KiTS23 and BraTS24, the benefits of adaptive clipping are more pronounced. These datasets exhibit substantial anatomical variability, class imbalance, and low-contrast regions, which exacerbate the impact of overly restrictive clipping under standard DP-FL. In such cases, ADP-FL substantially improves the segmentation of clinically critical subregions, reduces patient-level performance variability, and stabilizes training dynamics relative to the noisy and often unstable optimization observed with fixed clipping thresholds. This stabilization is reflected in both improved convergence profiles and more coherent qualitative predictions. Sensitivity analyses further indicate that ADP-FL is largely robust to the choice of clipping percentile for simpler tasks, and only moderately sensitive for more challenging multi-class segmentation problems, where higher percentiles better preserve task-relevant gradient information. Overall, these results suggest that adaptive clipping, when combined with dynamic noise injection, provides an effective mechanism for mitigating the adverse effects of differential privacy on optimization. By improving robustness, reducing variance across patients, and narrowing the performance gap with NP-FL, ADP-FL emerges as a reliable and scalable framework for privacy-preserving medical image segmentation across a wide range of modalities and clinical tasks.

\section{Conclusion}

This study introduces and validates an Adaptive Differentially Private Federated Learning (ADP-FL) framework designed to bridge the gap between rigorous patient privacy and high-fidelity medical image segmentation. Our results across three distinct modalities: 2D dermatoscopic images, 3D CT scans, and multi-parametric MRI, demonstrate that ADP-FL effectively mitigates the utility loss inherent in standard DP-FL. By dynamically adjusting clipping thresholds to the gradient distribution of clinical data, the framework achieves a near-optimal balance, tracking the performance of non-private models while operating under strict privacy guarantees. The practical implications of this work are twofold. First, ADP-FL addresses the "utility wall" that often prevents the adoption of differential privacy in clinical settings, particularly for complex 3D tasks like multi-subregion brain tumor segmentation where signal preservation is critical. Second, the framework’s robustness across diverse datasets with minimal hyperparameter tuning suggests a high degree of "out-of-the-box" readiness for real-world medical AI orchestration. By enabling decentralized training without exposing sensitive raw data, ADP-FL provides a viable pathway for multi-institutional collaboration that complies with stringent global data governance regulations. Despite these contributions, future research is warranted to evaluate ADP-FL within authentic, heterogeneous multi-institutional deployments to further assess cross-site generalization. Additionally, while this work focused on segmentation, exploring the framework’s efficacy in classification and longitudinal disease tracking could expand its clinical utility. In conclusion, ADP-FL represents a scalable, low-overhead solution for privacy-preserving medical analysis, ensuring that the advancement of diagnostic AI does not come at the cost of patient confidentiality.

\section*{Acknowledgments}

The author gratefully acknowledges the support of the University of Guelph, along with the scholarships, research grants, and valuable guidance from colleagues and mentors that made this work possible.

\bibliography{ref} 
\bibliographystyle{spiebib} 

\end{document}